\documentclass[9pt,twocolumn,twoside]{opticajnl}

\journal{optica} 

\setboolean{shortarticle}{false}

\title{Simultaneously sorting vector vortex beams of 120 modes}

\author[1]{Qi Jia}
\author[1]{Yanxia Zhang}
\author[1]{Bojian Shi}
\author[1]{Hang Li}
\author[1]{Xiaoxin Li}
\author[1]{Rui Feng}
\author[1]{Fangkui Sun}
\author[1,*]{Yongyin Cao}
\author[2,*]{Jian Wang}
\author[3,*]{Cheng-Wei Qiu}
\author[1,4,*]{Weiqiang Ding}

\affil[1]{Institute of Advanced Photonics, School of Physics, Harbin Institute of Technology, Harbin 150001, China}
\affil[2]{School of Physics, Harbin Institute of Technology, Harbin 150001, China}
\affil[3]{Department of Electrical and Computer Engineering, National University of Singapore, Singapore 117583, Singapore}
\affil[4]{Collaborative Innovation Center of Extreme Optics, Shanxi University, Taiyuan 030006, Shanxi, China}	

\affil[*]{Corresponding authors: 
\href{mailto:yycao@hit.edu.cn}{yycao@hit.edu.cn}; 
\href{mailto:hitwj@hit.edu.cn}{hitwj@hit.edu.cn}; \href{mailto:chengwei.qiu@nus.edu.sg}{chengwei.qiu@nus.edu.sg};  \href{mailto:wqding@hit.edu.cn}{wqding@hit.edu.cn}}

\begin{abstract}
  Polarization (P), angular index ($l$), and radius index ($p$) are three independent degrees of freedom (DoFs) of vector vortex beams, which have been widely used in optical communications, quantum optics, and information processing, etc. Although the sorting of one DoF can be achieved efficiently, it is still a great challenge to sort all these DoFs simultaneously in a compact and efficient way. Here, we propose a beam sorter to deal with all these three DoFs simultaneously by using a diffractive deep neural network (D$^2$NN), and experimentally demonstrated the robust sorting of $120$ Laguerre-Gaussian (LG) modes using a compact D$^2$NN formed by one spatial light modulator and one mirror only. The proposed beam sorter demonstrates the great potential of D$^2$NN in optical field manipulation and will benefit the diverse applications of vector vortex beams.
\end{abstract}
\setboolean{displaycopyright}{true}
\begin{document}
\maketitle

\section{Introduction}
Since 1992, Allen \textit{et al.} discovered that a light beam with the helical phase structured of $\exp (i l \varphi)$ carries an orbital angular momentum (OAM) of $l \hbar$ per photon ($l$, named the topological charge, which is usually an integer number) \cite{allenOrbitalAngularMomentum1992}. Unlike spin angular momentum (SAM), OAM, in principle, has infinite orthogonal eigenstates indicated by the integer $l$. OAM beams have been widely adopted in various applications with unprecedented performances due to the merit of the helical phase structure, including optical trapping \cite{ashkinAccelerationTrappingParticles1970,leeGiantColloidalDiffusivity2006,padgettTweezersTwist2011,huangSpintoorbitalAngularMomentum2021}, resolution enhanced microscopy \cite{tanHighresolutionWidefieldStandingwave2010,weiSub100nmResolutionPSIM2015,kozawaSuperresolutionImagingSuperoscillation2018}, nonlinear optics  \cite{gariepyCreatingHighHarmonicBeams2014,keren-zurNonlinearBeamShaping2016}, high-dimensional quantum states \cite{mairEntanglementOrbitalAngular2001,ficklerQuantumEntanglementHigh2012,wangDeterministicGenerationLargescale2022,wangFourdimensionalOrbitalAngular2022}, and high-capacity optical communications \cite{wangTerabitFreespaceData2012,bozinovicTerabitScaleOrbitalAngular2013,krennCommunicationSpatiallyModulated2014,leiMassiveIndividualOrbital2015,krennTwistedLightTransmission2016,laveryFreespacePropagationHighdimensional2017,jinPhyllotaxisinspiredNanosievesMultiplexed2021}. As the most representative OAM mode, the Laguerre-Gaussian (LG) mode has not only the topological charge $l$, but also the radius index of $p$ (a non-negative integer), which is also very important both in theory and application  \cite{karimiRadialQuantumNumber2014,plickPhysicalMeaningRadial2015,chenRealizationEinsteinPodolskyRosenParadox2019,wangDeterministicGenerationLargescale2022,zhangExperimentalInvestigationUncertainty2022}. As a vector field, except for $l$ and $p$, the polarization $P$ of a light beam is also a vital freedom, which cannot be ignored in practice. 
  
The sorting of vector vortex beams (VVBs) according to the indexes of $l$, $p$, and $P$ becomes a prerequisite operation in the various applications mentioned above. When only one of the indexes is considered, it is not difficult to sort. For example, the polarization $P$ can be separated efficiently using a polarization beam splitter (PBS). Recently, the sorting of angular index $l$ has attracted much attention, such as using optical geometric transformation \cite{berkhoutEfficientSortingOrbital2010,dudleyEfficientSortingBessel2013,wenSpiralTransformationHighResolution2018,wenCompactHighperformanceVortex2020,guoSpindecoupledMetasurfaceSimultaneous2021,chengUltracompactOrbitalAngular2022}, which only uses less than two phase modulators. It is worth noting that using metasurface can sort both the polarization $P$ and angular index $l$ simultaneously \cite{guoSpindecoupledMetasurfaceSimultaneous2021}. However, it is not satisfactory for the radius index $p$, because $p$ is mainly reflected in the intensity. The Gouy phase of LG mode is related to $l$, and $p$, which provides a potential method to sort $l$ and $p$ simultaneously \cite{zhouSortingPhotonsRadial2017,guGouyPhaseRadial2018,weiActiveSortingOrbital2020}. Unfortunately, however, the Gouy phase is degenerate for the angular index of $\pm l$, which means this method can not sort $l$ and $-l$ directly. Alternatively, using the combination of Dove prisms and PBSs can sort $l$ also\cite{zhangMimickingFaradayRotation2014}, which generally requires a system cascaded by many units. This means that the system's complexity increases rapidly with its sorting capability, making it unsuitable for a larger mode number.
Therefore, up to now, it is still a great challenge to sort all three indexes with high efficiency, low crosstalk, and a compact system.

In such a circumstance, based on the diffractive deep neural network (D$^2$NN), we theoretically propose a vector vortex beam sorter and experimentally demonstrate the sorting of $120$ modes (i.e., $l$ from $-7$ to $7$, $p$ from $0$ to $3$, and $P$ for $\left|\mathrm{H}\right\rangle$ and $\left|\mathrm{V}\right\rangle$). Our key idea is to transform a polarized diffractive deep neural network into two scalar diffractive deep neural networks. First, we use a calcite crystal to split the orthogonal polarization ($\left|\mathrm{H}\right\rangle$ and $\left|\mathrm{V}\right\rangle$). And then, after the half-wave plate, $\left|\mathrm{V}\right\rangle$ changes into $\left|\mathrm{H}\right\rangle$. Last, we use two 5-layer D$^2$NNs consisting of a spatial light modulator and mirror to sort LG modes from two channels. We demonstrated the output results of several vector structured beams, tested the robustness in atmospheric turbulence, and obtained satisfactory results in the experiment. We believe this sorter benefits the applications of vector structured beams. 

\section{Methods}

In 2018, Lin \textit{et al.} proposed the diffractive deep neural network (D$^2$NN) \cite[]{linAllopticalMachineLearning2018}. After that, D$^2$NN has made great progress in image processing \cite[]{liClassspecificDifferentialDetection2019,yanFourierspaceDiffractiveDeep2019,menguScaleShiftRotationInvariant2021,zhouLargescaleNeuromorphicOptoelectronic2021,rahmanEnsembleLearningDiffractive2021,menguClassificationReconstructionSpatially2022,luoComputationalImagingComputer2022,baiImageNotImage2022,menguAllOpticalPhase2022,xuMultichannelOpticalComputing2022}. Besides, due to the photon owned more degrees of freedom (i.e., polarization, wavelength, mode), these DoFs provide a new way of thinking \cite[]{luoDesignTaskspecificOptical2019,fontaineLaguerreGaussianModeSorter2019,brandtHighdimensionalQuantumGates2020,veliTerahertzPulseShaping2021,fangPerformanceOptimizationMultiplane2021,hiekkamakiHighDimensionalTwoPhotonInterference2021,liPolarizationMultiplexedDiffractive2022,weiParallelArrayedWaveguide2022,jiaCompensatingDistortedOAM2022,luoMetasurfaceenabledOnchipMultiplexed2022,jiaUniversalTranslationOperator2022}. Normally, D$^2$NN theoretically has the ability to process polarization information. However, corresponding polarization control units are required in practice. Using metasurfaces seems like a good option, which modulates both polarizations and enables a system-on-chip \cite[]{luoMetasurfaceenabledOnchipMultiplexed2022,chengUltracompactOrbitalAngular2022}. Unfortunately, in the experiment, it is a huge challenge to align the adjacent layer, and the transmittance is also regrettable for multi-metasurface in the visible spectrum. Fortunately, we have a more feasible method for vector structured beams using the spatial light modulator to achieve this sorter. Without loss of generality, the vector structured beam can be described as, 
\begin{equation}
    \label{eq1}
    \left| \psi \right\rangle = \sum_{l, p} \left( \alpha_{l}^{p} \left| \mathrm{LG}_{l}^{p} \right\rangle \left| \mathrm{H}\right\rangle + \beta_{l}^{p} \left| \mathrm{LG}_{l}^{p} \right\rangle \left| \mathrm{V} \right\rangle \right),
\end{equation}
\noindent where $\alpha_{l}^{p}$ and $\beta_{l}^{p}$ are normalization coefficient. $ \left| \mathrm{H} \right\rangle $ and $ \left| \mathrm{V} \right\rangle $ mean horizontal polarization and vertical polarization, respectively. $ \left| \mathrm{LG}_{l}^{p} \right\rangle $ is the Laguerre-Gaussian mode with circular symmetry, which are the earliest reported vortex beams carrying OAM and described as \cite[]{allenOrbitalAngularMomentum1992}, 

  \begin{equation}
    \begin{aligned}
    \mathrm{LG}_{l}^{p}(r, \varphi, z)=& \frac{{C_{l}^{p}}}{\omega(z)}\left(\frac{r \sqrt{2}}{\omega(z)}\right)^{|l|} L_p^{|l|} \left(\frac{2 r^2}{\omega(z)^2}\right) \\
    & \times \exp \left(\frac{-r^2}{\omega(z)^2}\right) \exp \left(\frac{-i k r^2 z}{2\left(z^2+z_R^2\right)}\right) \\
    & \times \exp (i l \varphi) \exp (i(2 p+|l|+1) \psi(z))
    \end{aligned},
  \end{equation}
  \noindent where ${C^{p}_{l}}$ is the normalization constant, $k$ is the wave number, $l$ is the angular quantum number, and $p$ is the radial quantum number. ${\omega(z)}$, $\psi(z)$ and $z_{R}$ are 

  \begin{equation}
    \begin{aligned}
    &\omega(z)=\omega_0 \sqrt{1+\left(\frac{z}{z_R}\right)^2} \\
    &\psi(z)=\arctan \frac{z}{z_R}\\
    &z_R = \pi \omega_{0}^{2}/\lambda
    \end{aligned}, 
  \end{equation}
  \noindent where $\lambda$ is the wavelength and $\omega_{0}$ is the basement membrane waist radius.

  Notoriously, the $ \left| \mathrm{H} \right\rangle $ and $ \left| \mathrm{V} \right\rangle $ are orthogonal. Thus, we can separate $ \left| \mathrm{H} \right\rangle $ and $ \left| \mathrm{V} \right\rangle $ first, which is easy to achieve by polarizing elements (i.e., polarizing beam splitter and calcite crystal). Then, we sort LG modes from two channels. This way, we transform the polarization diffractive deep neural network into two scalar diffractive deep neural networks. 
  
  For the scalar diffractive deep neural network, according to the angular spectrum theory, the propagation of an optical field can be described as \cite[]{goodmanIntroductionFourierOptics2005a}
  \begin{equation}
    E(x, y,z)=\hat{\mathcal{F} }^{-1} \exp \left(i z \sqrt{k^2-k_x^2-k_y^2}\right) \hat{\mathcal{F} } E(x, y, 0), 
  \end{equation}
  \noindent where$\hat{\mathcal{F} }$ is the operator for the 2D Fourier transform, and correspondingly, $\hat{\mathcal{F} }^{-1}$ represents the inverse 2D Fourier transform. Here, $k$ is the angular wavenumber of the fields, $k_{x}$ and $k_{y}$ are spatial frequencies. Moreover, the backward propagating and gradient descent algorithms are used to train the diffractive deep neural networks (more details shown in Supplement). 

  \section{Results}  
\begin{figure}[ht!]
    \centering
    \includegraphics[width = 8.5cm]{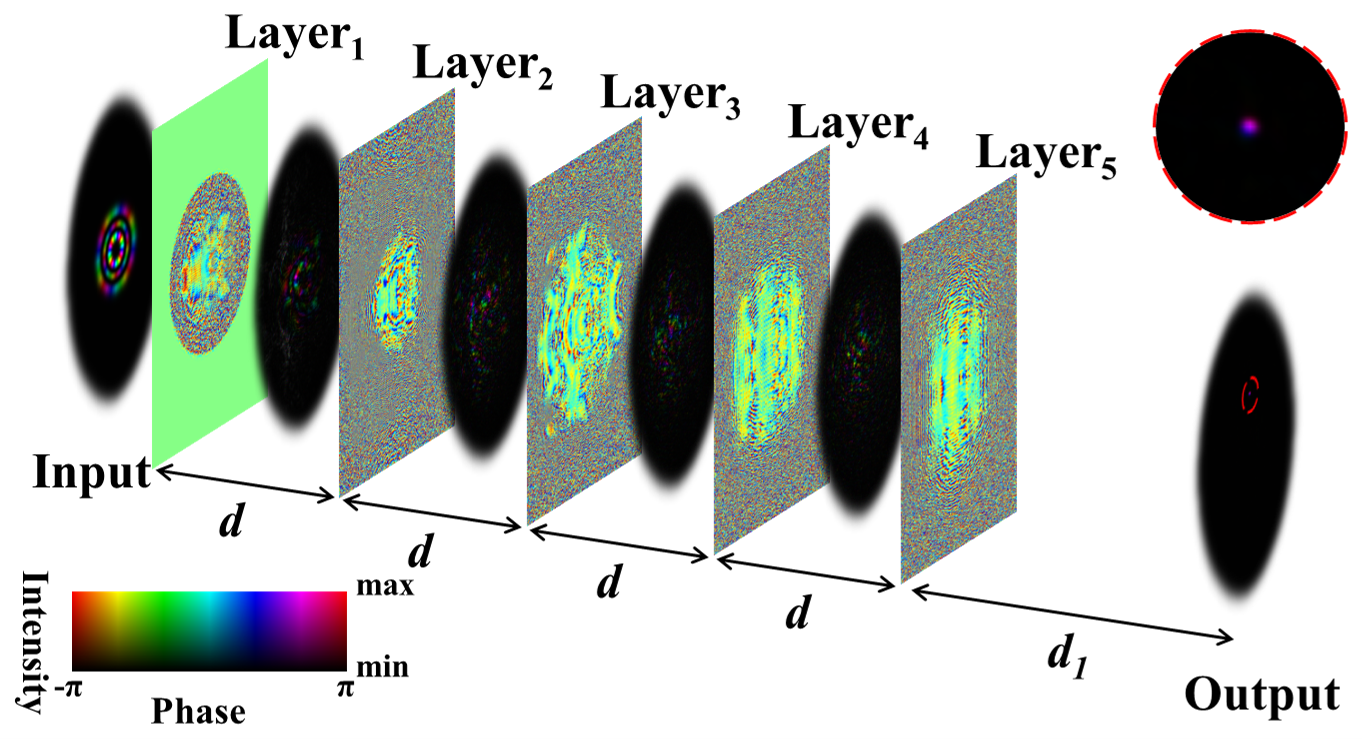}
    \caption{
      Schematic for the sorting of LG beams with angular quantum index $l$ and radius quantum index $p$ realized by $5$ phase planes of $\mathrm{Layer}_{1,2,3,4,5}$, the distance of each layer $d$ is $3.51$ cm, and the distance from the last layer to the output plane $d_{1}$ is $9.3$ cm. For clarity, the red round shows the enlarged view of the output beam.
    }
    \label{img1}
  \end{figure}

  For the vector structured beam sorter, we first choose the LG modes and Gaussian spots as the input train set and the target output train set of the LG mode sorter. For the train LG set, angular quantum index $l$ is from $-7$ to $7$, and radius quantum index $p$ is from $0$ to $3$ (total of $60$ modes). For D$^2$NN, there are a total of $5$ neural layers, and each layer has $310 \times 420$ optical neurons with the neuron size of $8 \times 8$ $\mu$m (matched the pixel size of SLM in the experiment), and the wavelength $\lambda$ is $532$ nm. Moreover, the distance of each layer $d$ is $3.51$ cm, and the distance from the last layer to the output plane $d_{1}$ is $9.3$ cm. After training, the result is shown in Fig. \ref{img1}.

  As shown in Fig. \ref{img1}, the D$^2$NN can sort the LG mode after training. Furthermore, we show the more detailed propagation of the $\left| \mathrm{LG}_{4}^{3} \right\rangle$, and the propagation distance of the adjacent frame is about $3.2$ mm in Video1. Apart from Video1, we also show the $\left| \psi \right\rangle = \left|\mathrm{LG}^{1}_{2}\right\rangle +\left| \mathrm{LG}^{3}_{-7} \right\rangle $ and $\left| \psi \right\rangle = \left|\mathrm{LG}^{2}_{5}\right\rangle +\left| \mathrm{LG}^{0}_{-2} \right\rangle +\left| \mathrm{LG}^{1}_{-7} \right\rangle $ in Video2 and Video3. These videos show that the D$^2$NN recognizes the feature of LG modes and outputs it at the corresponding position with low crosstalk in simulation. For the VSB sorter, we use two LG mode sorters to process LG modes from two polarization channels $\left| \mathrm{H} \right\rangle$ and $\left| \mathrm{V} \right\rangle$.

	\begin{figure*}[ht!]
		\centering\includegraphics[width = \linewidth]{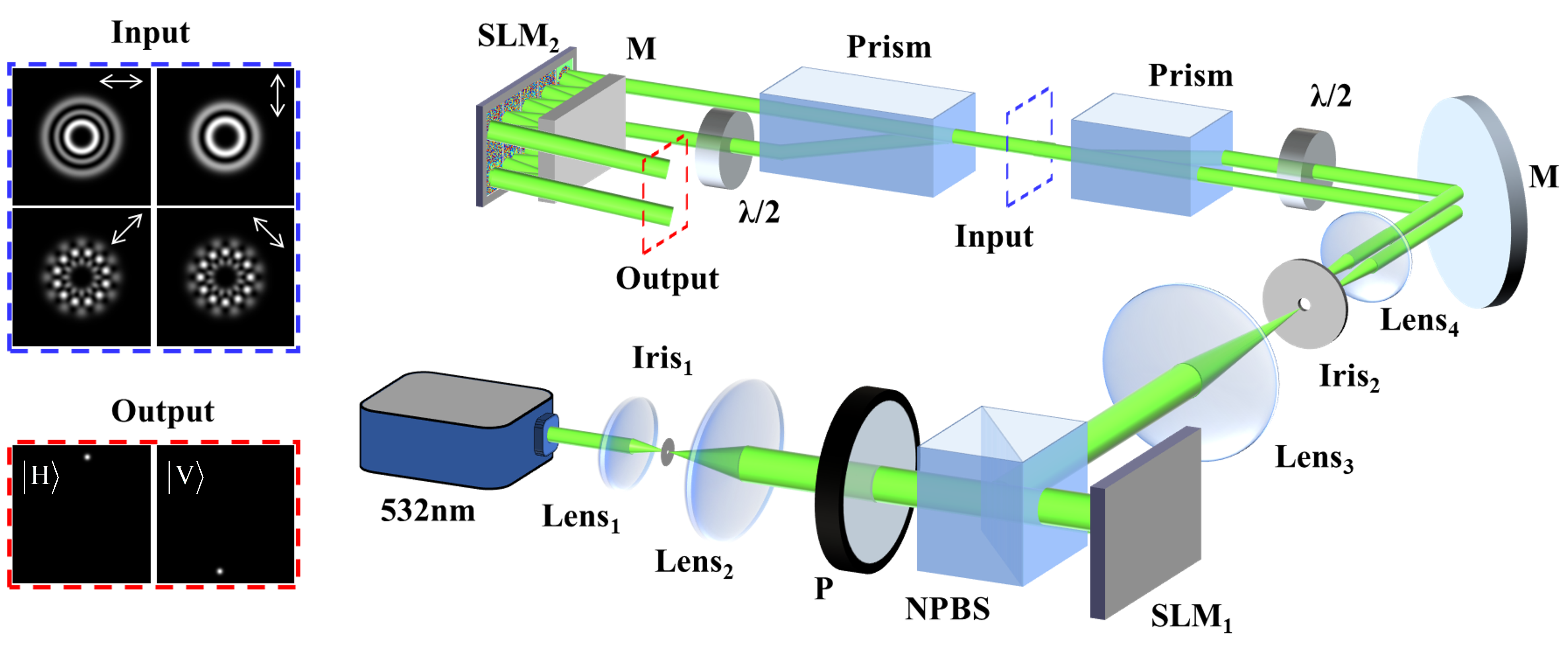}
		\caption{
			Experimental setup of vector structured beams sorting using the D$^2$NN. The Gaussian laser beam with a wavelength of $532$ nm is expanded and collimated by Lens$_1$ and Lens$_2$ and followed by a polarizer P. Then SLM$_1$ converts the Gaussian beam into two beams, corresponding to two orthogonal polarization states. After 4-\textit{f} system, using HWP changes the polarization of one beam from $\left|\mathrm{H} \right\rangle$ to $\left|\mathrm{V} \right\rangle$. Then using a calcite crystal, combines two orthogonal polarization beams into one. Last, the input vector structured beams are sorted by D$^2$NN, consisting of a calcite crystal, HWP, SLM$_2$, and M. In the blue dotted line, the input vector structured beam is measured after the linear polarizer aligned at $0^{\circ}$, $90^{\circ}$, $45^{\circ}$, and $135^{\circ}$. In the red dotted line, the output result is the channel $\left|\mathrm{H} \right\rangle$ and $\left|\mathrm{V} \right\rangle$ successively. List of abbreviations: non-polarizing beam splitter (NPBS); spatial light modulator (SLM); half-wave plate (HWP); mirror (M); and polarizer (P).
		}\label{img2}
	\end{figure*}

  In order to verify the performance of the vector structured beam sorter in practice, we built a corresponding experimental system. Fig. \ref{img2} shows the schematic setup of the VSB sorter used in our experiment. The incident Gaussian laser beam with a wavelength of 532 nm is expanded and collimated by Lens$_1$, Lens$_2$, and Iris$_1$. Followed by the polarizer P, the polarization of the incident beam is changed into $\left| \mathrm{H} \right\rangle$. Then, two forked phase hologram is loaded on SLM$_1$ (Holoeye PLUTO-2) to generate two exiting LG beams with various $l$ and $p$, after the pinhole filtering (Iris$_2$) and 4-\textit{f} system (Lens$_3$ and Lens$_4$). Through the HWP, we transformed the polarization of one of the beams from $\left| \mathrm{H} \right\rangle$ to $\left| \mathrm{V} \right\rangle$. Finally, we combined the two orthogonally polarized LG beams into one using a calcite crystal and obtained the corresponding VSB. As we pointed out earlier, we use a calcite crystal and HWP to separate the two orthogonal polarization ($\left| \mathrm{H} \right\rangle$, $\left| \mathrm{V} \right\rangle$) and transform $\left| \mathrm{V} \right\rangle$ into $\left| \mathrm{H} \right\rangle$, corresponding to the upper and lower D$^2$NNs. Here, transforming $\left| \mathrm{V} \right\rangle$ into $\left| \mathrm{H} \right\rangle$ is because the SLM only modulator the $\left| \mathrm{H} \right\rangle$. Then, we use SLM$_2$ and a plane mirror M to compose two 5-layer D$^2$NNs by dividing SLM$_2$ into $10$ regions (as shown in  Supplement). The numerically solved phase patterns are loaded in the $5$ upper parts of SLM$_2$. Moreover, the $5$ lower parts of SLM$_2$ are the same. Using this configuration, one SLM is enough to realize the two 5-layer D$^2$NNs shown in Fig. \ref{img1}, which is very convenient in the experiment.  

\begin{figure*}[ht!]
    \centering\includegraphics[width=\linewidth]{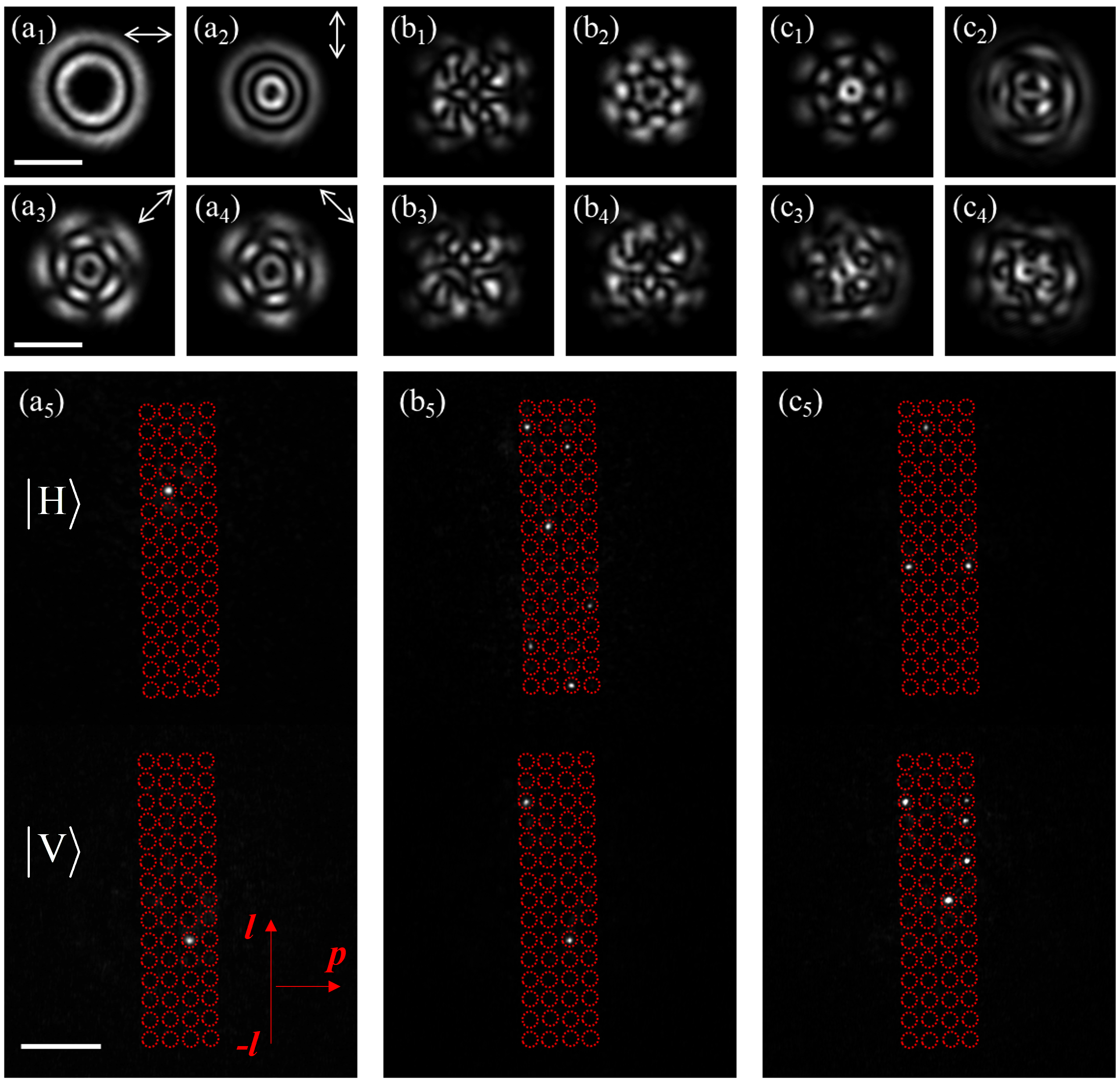}
    \caption{ \label{img3}
      Experimental results of the vector structured sorter for different input beams. (a$_1$-a$_5$) the input beam and output beam are measured in the experiment, and the input beam is $\left|\psi_{in}\right\rangle = \left|\mathrm{LG}_{3}^{1}\right\rangle  \left|\mathrm{H} \right\rangle + \left|\mathrm{LG}_{-2}^{2}\right\rangle  \left|\mathrm{V} \right\rangle$. (a$_1$-a$_4$) are the input beam measured after the linear polarizer aligned at $0^{\circ}$, $90^{\circ}$, $45^{\circ}$, and $135^{\circ}$. (a$_5$) is the output result, in which the upper half is the $\left|\mathrm{H} \right\rangle $ channel, and the lower half is the $\left|\mathrm{V} \right\rangle $ channel (For each channel, $l$ is from $-7$ to $7$ and $p$ is from $0$ to $3$). (b$_1$-b$_5$) are the same as (a$_1$-a$_5$), except for $\left|\psi_{in}\right\rangle =\left( \left|\mathrm{LG}_{6}^{0}\right\rangle +\left|\mathrm{LG}_{5}^{2}\right\rangle +\left|\mathrm{LG}_{1}^{1}\right\rangle+\left|\mathrm{LG}_{-3}^{3}\right\rangle+\left|\mathrm{LG}_{-5}^{0}\right\rangle+\left|\mathrm{LG}_{-7}^{2}\right\rangle \right)  \left|\mathrm{H} \right\rangle + \left( \left|\mathrm{LG}_{5}^{0}\right\rangle +\left|\mathrm{LG}_{2}^{2}\right\rangle \right)  \left|\mathrm{V} \right\rangle$. (c$_1$-c$_5$) are the same as (a$_1$-a$_5$), except for $\left|\psi_{in}\right\rangle = \left( \left|\mathrm{LG}_{6}^{1}\right\rangle +\left|\mathrm{LG}_{-1}^{0}\right\rangle +\left|\mathrm{LG}_{-1}^{3}\right\rangle \right)  \left|\mathrm{H} \right\rangle + \left( \left|\mathrm{LG}_{5}^{0}\right\rangle +\left|\mathrm{LG}_{5}^{3}\right\rangle +\left|\mathrm{LG}_{4}^{3}\right\rangle+\left|\mathrm{LG}_{2}^{3}\right\rangle+\left|\mathrm{LG}_{0}^{2}\right\rangle \right)  \left|\mathrm{V} \right\rangle $. (Scale bar: 500 $\mu$m).
    }
  \end{figure*}

  Fig. \ref{img3} shows this system's vector structured beam sorting experimental results. In the upper two lines of Fig. \ref{img3}, we show the input VSBs measured after the linear polarizer aligned at $0^{\circ}$, $90^{\circ}$, $45^{\circ}$, and $135^{\circ}$. Furthermore, in the lower line, we show the sorting results. The output channels of polarization are the $\left| \mathrm{H} \right\rangle$ and $\left| \mathrm{V} \right\rangle$ channels from top to bottom. Furthermore, for the output channels of LG mode, the channels of angular quantum index $l$ (from bottom to top) are from $-7$ to $7$, and the channels of radial quantum index $p$ (from left to right) are from $0$ to $3$. In Figs. \ref{img3}. (a$_1$-a$_5$), we show the input and output result of VSB $\left|\psi_{in}\right\rangle = \left|\mathrm{LG}_{3}^{1}\right\rangle  \left|\mathrm{H} \right\rangle + \left|\mathrm{LG}_{-2}^{2}\right\rangle  \left|\mathrm{V} \right\rangle$. The vector structured beam sorter can separate the different VSBs in the designed positions, and the crosstalk of different channels is low. Moreover, we also test the system's performances for incident multi-mode (up to $8$ modes) and get satisfactory results, as shown in Figures. \ref{img3}. (b$_1$-b$_5$ \& c$_1$-c$_5$).  
  
\section{Discussion}

\begin{figure}[ht!]
    \centering\includegraphics[width=\linewidth]{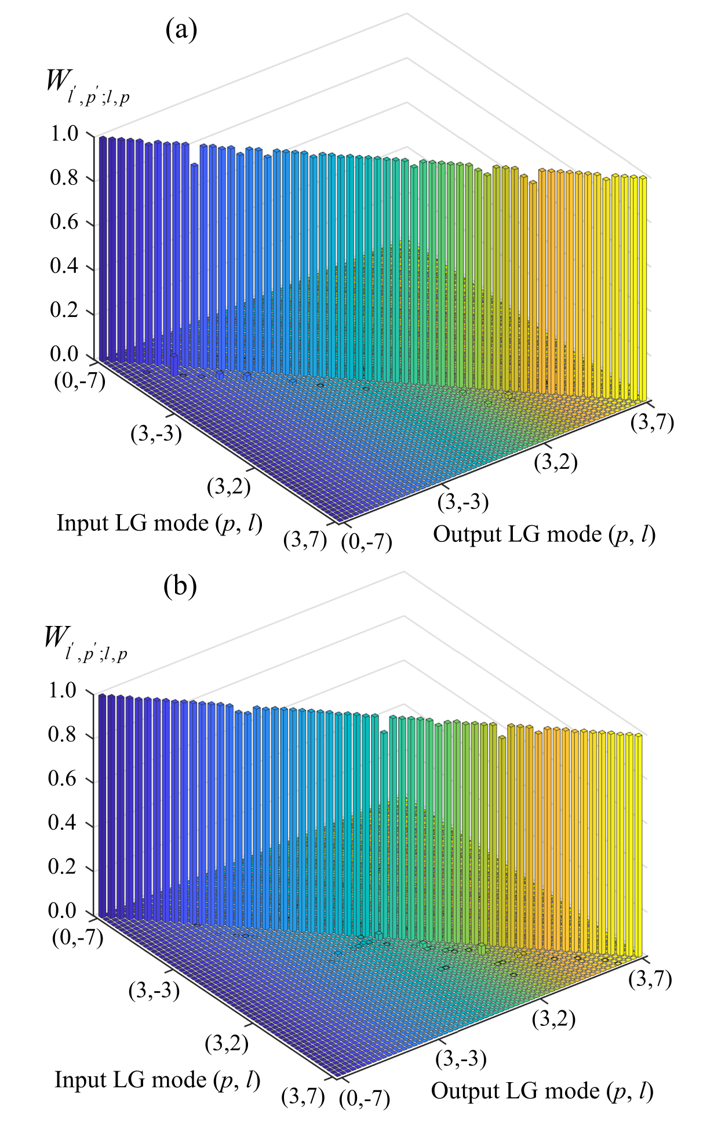}
    \caption{ \label{img4}
      Normalized energy $W_{{l^{\prime}},{p^{\prime}};l,p}$ in the sorting of the $60$ modes for each position, which is the measured energy of channel $\mathrm{LG}_{l^{\prime}}^{p^{\prime}}$mode normalized by the total energy of output channels for incident mode $\mathrm{LG}_{l}^{p}$. Here $l$ and ${l^{\prime}}$ change from $-7$ to $7$, and $p$ and ${p^{\prime}}$ change from $0$ to $3$. All the parameters are the same as those in FIG. \ref{img3}. (a) is measured in $\left|\mathrm{H} \right\rangle $ polarization. (b) is measured in $\left|\mathrm{V} \right\rangle $ polarization. 
    }
  \end{figure}

  Besides multi-mode results, the crosstalks between different channels are also critical in practice. Compared with the forked grating and Dammann grating, our method only gets the output quasi-Gaussian beam without other outputs. In Fig. \ref{img4}, we show all the normalized energies $W_{l^{\prime},p^{\prime};l,p}$ from two polarizations, which is the measured energy of LG$_{l'}^{p'}$ mode normalized by the incident energy of LG$_l^p$ mode. All the other parameters used in Fig. \ref{img4} are the same as those in Fig. \ref{img3}. Fig. \ref{img4} shows that the energy efficiencies of the sorting are high, and the average value of $E_{l,p;l,p}$ is about $99.43\%$.  The average crosstalk of $W_{l^{\prime},p^{\prime};l,p}$ with $l\neq l^{\prime}$ and $p\neq p^{\prime}$ is near $0\%$. The maximum crosstalk does not exceed $-12$dB. It is important for the applications of VBS. It is noted that the crosstalk is mainly due to the aberrations accumulated when passing through the SLMs. In principle, these errors can be compensated using several methods or the Zernike function to generate the compensated phase \cite{menguScaleShiftRotationInvariant2021,zhouLargescaleNeuromorphicOptoelectronic2021}.

  In practice, it is a challenge to sort the distorted LG beams \cite[]{renAdaptiveopticsbasedSimultaneousPre2014,fuNonprobeCompensationOptical2017}.Here, we use the distorted VBSs to test the sorter's accuracy (more details shown in Supplement). Compared with ordinary VSB, the crosstalk of the system affects its use. The system's error rate is above $23\%$, and most errors occur in $l$. For $p$, the error rate is about $5\%$. Due to atmospheric turbulence, the intensity and phase have been distorted, which changes the mode of VBS. Although the system cannot obtain satisfactory results for distorted beams, which can use adaptive optics to compensate for the distorted wavefront \cite[]{renAdaptiveopticsbasedSimultaneousPre2014,fuNonprobeCompensationOptical2017}.

	We presented a vector structured beam sorter based on the diffractive deep neural network (120 modes in total). Compared with optical geometric transformation, our method can simultaneously sort polarization P, the angular quantum index $l$, and the radius quantum index $p$. The vector structured beam sorter can be separated into two LG mode sorters. After training, D$^2$NN  can recognize the feature of LG modes ($l$ and $p$) and output it at the corresponding position with low crosstalk in simulation. To achieve this sorter in the experiment, we firstly use a polarizing splitter system to separate two orthogonal polarizations ($\left|\mathrm{V} \right\rangle$, $\left| \mathrm{V} \right\rangle$) and change $\left|\mathrm{V} \right\rangle$ to $\left| \mathrm{H} \right\rangle$, consisting of a calcite crystal and HWP. Furthermore, we use a spatial light modulator and mirror to form two 5-layer D$^2$NNs in the visible spectrum of $532$ nm. Moreover, the crosstalk of the system is less than $-12$dB. We believe our vector structural beam sorter would because of a beneficial component in practical multi-dimensional multiplexed optical communication and high-dimensional quantum information systems.

\begin{backmatter}
\bmsection{Funding} 
National Natural Science Foundation of China (Grant Nos. 11874134, 12274105),  Heilongjiang Natural Science Funds for Distinguished Young Scholar (Grant No. JQ2022A001). Fundamental Research Funds for the Central Universities (HIT.OCEF.2021020).

\bmsection{Acknowledgments} 
We acknowledge the HPC Studio at the School of Physics, Harbin Institute of Technology, for access to computing resources through INSPURHPC@PHY.HIT.

\bmsection{Disclosures} The authors declare no conflicts of interest.

\bmsection{Data Availability Statement} Data underlying the results presented in this paper are not publicly available at this time but may be obtained from the authors upon reasonable request.

\bmsection{Supplemental document}
See Supplement for supporting content. 

\end{backmatter}


\begin{thebibliography}{10}
		\newcommand{\enquote}[1]{``#1''}
		
		\bibitem{allenOrbitalAngularMomentum1992}
		L.~Allen, M.~W. Beijersbergen, R.~J.~C. Spreeuw, and J.~P. Woerdman,
		\enquote{Orbital angular momentum of light and the transformation of
			laguerre-gaussian laser modes,} {\protect\JournalTitle{Physical Review A}}
		\textbf{45}, 8185--8189 (1992).
		
		\bibitem{ashkinAccelerationTrappingParticles1970}
		A.~Ashkin, \enquote{Acceleration and trapping of particles by radiation
			pressure,} {\protect\JournalTitle{Physical Review Letters}} \textbf{24},
		156--159 (1970).
		
		\bibitem{leeGiantColloidalDiffusivity2006}
		S.-H. Lee and D.~G. Grier, \enquote{Giant colloidal diffusivity on corrugated
			optical vortices,} {\protect\JournalTitle{Physical Review Letters}}
		\textbf{96}, 190601 (2006).
		
		\bibitem{padgettTweezersTwist2011}
		M.~Padgett and R.~Bowman, \enquote{Tweezers with a twist,}
		{\protect\JournalTitle{Nature Photonics}} \textbf{5}, 343--348 (2011).
		
		\bibitem{huangSpintoorbitalAngularMomentum2021}
		S.-Y. Huang, G.-L. Zhang, Q.~Wang, M.~Wang, C.~Tu, Y.~Li, and H.-T. Wang,
		\enquote{Spin-to-orbital angular momentum conversion via light intensity
			gradient,} {\protect\JournalTitle{Optica}} \textbf{8}, 1231--1236 (2021).
		
		\bibitem{tanHighresolutionWidefieldStandingwave2010}
		P.~S. Tan, X.-C. Yuan, G.~H. Yuan, and Q.~Wang, \enquote{High-resolution
			wide-field standing-wave surface plasmon resonance fluorescence microscopy
			with optical vortices,} {\protect\JournalTitle{Applied Physics Letters}}
		\textbf{97}, 241109 (2010).
		
		\bibitem{weiSub100nmResolutionPSIM2015}
		S.~Wei, T.~Lei, L.~Du, C.~Zhang, H.~Chen, Y.~Yang, S.~W. Zhu, and X.-C. Yuan,
		\enquote{Sub-100nm resolution psim by utilizing modified optical vortices
			with fractional topological charges for precise phase shifting,}
		{\protect\JournalTitle{Optics Express}} \textbf{23}, 30143--30148 (2015).
		
		\bibitem{kozawaSuperresolutionImagingSuperoscillation2018}
		Y.~Kozawa, D.~Matsunaga, and S.~Sato, \enquote{Superresolution imaging via
			superoscillation focusing of a radially polarized beam,}
		{\protect\JournalTitle{Optica}} \textbf{5}, 86--92 (2018).
		
		\bibitem{gariepyCreatingHighHarmonicBeams2014}
		G.~Gariepy, J.~Leach, K.~T. Kim, T.~J. Hammond, E.~Frumker, R.~W. Boyd, and
		P.~B. Corkum, \enquote{Creating high-harmonic beams with controlled orbital
			angular momentum,} {\protect\JournalTitle{Physical Review Letters}}
		\textbf{113}, 153901 (2014).
		
		\bibitem{keren-zurNonlinearBeamShaping2016}
		S.~{Keren-Zur}, O.~Avayu, L.~Michaeli, and T.~Ellenbogen, \enquote{Nonlinear
			beam shaping with plasmonic metasurfaces,} {\protect\JournalTitle{ACS
				Photonics}} \textbf{3}, 117--123 (2016).
		
		\bibitem{mairEntanglementOrbitalAngular2001}
		A.~Mair, A.~Vaziri, G.~Weihs, and A.~Zeilinger, \enquote{Entanglement of the
			orbital angular momentum states of photons,} {\protect\JournalTitle{Nature}}
		\textbf{412}, 313--316 (2001).
		
		\bibitem{ficklerQuantumEntanglementHigh2012}
		R.~Fickler, R.~Lapkiewicz, W.~N. Plick, M.~Krenn, C.~Schaeff, S.~Ramelow, and
		A.~Zeilinger, \enquote{Quantum entanglement of high angular momenta,}
		{\protect\JournalTitle{Science}} \textbf{338}, 640--643 (2012).
		
		\bibitem{wangDeterministicGenerationLargescale2022}
		X.~Wang, S.~Yu, S.~Liu, K.~Zhang, Y.~Lou, W.~Wang, and J.~Jing,
		\enquote{Deterministic generation of large-scale hyperentanglement in three
			degrees of freedom,} {\protect\JournalTitle{Advanced Photonics Nexus}}
		\textbf{1}, 016002 (2022).
		
		\bibitem{wangFourdimensionalOrbitalAngular2022}
		C.~Wang, Y.~Chen, and L.~Chen, \enquote{Four-dimensional orbital angular
			momentum bell-state measurement assisted by the auxiliary polarization and
			path degrees of freedom,} {\protect\JournalTitle{Optics Express}}
		\textbf{30}, 34468--34478 (2022).
		
		\bibitem{wangTerabitFreespaceData2012}
		J.~Wang, J.-Y. Yang, I.~M. Fazal, N.~Ahmed, Y.~Yan, H.~Huang, Y.~Ren, Y.~Yue,
		S.~Dolinar, M.~Tur, and A.~E. Willner, \enquote{Terabit free-space data
			transmission employing orbital angular momentum multiplexing,}
		{\protect\JournalTitle{Nature Photonics}} \textbf{6}, 488--496 (2012).
		
		\bibitem{bozinovicTerabitScaleOrbitalAngular2013}
		N.~Bozinovic, Y.~Yue, Y.~Ren, M.~Tur, P.~Kristensen, H.~Huang, A.~E. Willner,
		and S.~Ramachandran, \enquote{Terabit-scale orbital angular momentum mode
			division multiplexing in fibers,} {\protect\JournalTitle{Science}}
		\textbf{340}, 1545--1548 (2013).
		
		\bibitem{krennCommunicationSpatiallyModulated2014}
		M.~Krenn, R.~Fickler, M.~Fink, J.~Handsteiner, M.~Malik, T.~Scheidl, R.~Ursin,
		and A.~Zeilinger, \enquote{Communication with spatially modulated light
			through turbulent air across vienna,} {\protect\JournalTitle{New Journal of
				Physics}} \textbf{16}, 113028 (2014).
		
		\bibitem{leiMassiveIndividualOrbital2015}
		T.~Lei, M.~Zhang, Y.~Li, P.~Jia, G.~N. Liu, X.~Xu, Z.~Li, C.~Min, J.~Lin,
		C.~Yu, H.~Niu, and X.~Yuan, \enquote{Massive individual orbital angular
			momentum channels for multiplexing enabled by dammann gratings,}
		{\protect\JournalTitle{Light: Science \& Applications}} \textbf{4},
		e257--e257 (2015).
		
		\bibitem{krennTwistedLightTransmission2016}
		M.~Krenn, J.~Handsteiner, M.~Fink, R.~Fickler, R.~Ursin, M.~Malik, and
		A.~Zeilinger, \enquote{Twisted light transmission over 143 km,}
		{\protect\JournalTitle{Proceedings of the National Academy of Sciences}}
		\textbf{113}, 13648--13653 (2016).
		
		\bibitem{laveryFreespacePropagationHighdimensional2017}
		M.~P.~J. Lavery, C.~Peuntinger, K.~G{\"u}nthner, P.~Banzer, D.~Elser, R.~W.
		Boyd, M.~J. Padgett, C.~Marquardt, and G.~Leuchs, \enquote{Free-space
			propagation of high-dimensional structured optical fields in an urban
			environment,} {\protect\JournalTitle{Science Advances}} \textbf{3}, e1700552
		(2017).
		
		\bibitem{jinPhyllotaxisinspiredNanosievesMultiplexed2021}
		Z.~Jin, D.~Janoschka, J.~Deng, L.~Ge, P.~Dreher, B.~Frank, G.~Hu, J.~Ni,
		Y.~Yang, J.~Li, C.~Yu, D.~Lei, G.~Li, S.~Xiao, S.~Mei, H.~Giessen, F.~M. {zu
			Heringdorf}, and C.-W. Qiu, \enquote{Phyllotaxis-inspired nanosieves with
			multiplexed orbital angular momentum,} {\protect\JournalTitle{eLight}}
		\textbf{1}, 5 (2021).
		
		\bibitem{karimiRadialQuantumNumber2014}
		E.~Karimi, R.~W. Boyd, P.~{de la Hoz}, H.~{de Guise}, J.~{\v R}eh{\'a}{\v c}ek,
		Z.~Hradil, A.~Aiello, G.~Leuchs, and L.~L. {S{\'a}nchez-Soto},
		\enquote{Radial quantum number of laguerre-gauss modes,}
		{\protect\JournalTitle{Physical Review A}} \textbf{89}, 063813 (2014).
		
		\bibitem{plickPhysicalMeaningRadial2015}
		W.~N. Plick and M.~Krenn, \enquote{Physical meaning of the radial index of
			laguerre-gauss beams,} {\protect\JournalTitle{Physical Review A}}
		\textbf{92}, 063841 (2015).
		
		\bibitem{chenRealizationEinsteinPodolskyRosenParadox2019}
		L.~Chen, T.~Ma, X.~Qiu, D.~Zhang, W.~Zhang, and R.~W. Boyd,
		\enquote{Realization of the einstein-podolsky-rosen paradox using radial
			position and radial momentum variables,} {\protect\JournalTitle{Physical
				Review Letters}} \textbf{123}, 060403 (2019).
		
		\bibitem{zhangExperimentalInvestigationUncertainty2022}
		Z.~Zhang, D.~Zhang, X.~Qiu, Y.~Chen, S.~{Franke-Arnold}, and L.~Chen,
		\enquote{Experimental investigation of the uncertainty principle for radial
			degrees of freedom,} {\protect\JournalTitle{Photonics Research}} \textbf{10},
		2223--2228 (2022).
		
		\bibitem{berkhoutEfficientSortingOrbital2010}
		G.~C.~G. Berkhout, M.~P.~J. Lavery, J.~Courtial, M.~W. Beijersbergen, and M.~J.
		Padgett, \enquote{Efficient sorting of orbital angular momentum states of
			light,} {\protect\JournalTitle{Physical Review Letters}} \textbf{105}, 153601
		(2010).
		
		\bibitem{dudleyEfficientSortingBessel2013}
		A.~Dudley, T.~Mhlanga, M.~Lavery, A.~McDonald, F.~S. Roux, M.~Padgett, and
		A.~Forbes, \enquote{Efficient sorting of bessel beams,}
		{\protect\JournalTitle{Optics Express}} \textbf{21}, 165--171 (2013).
		
		\bibitem{wenSpiralTransformationHighResolution2018}
		Y.~Wen, I.~Chremmos, Y.~Chen, J.~Zhu, Y.~Zhang, and S.~Yu, \enquote{Spiral
			transformation for high-resolution and efficient sorting of optical vortex
			modes,} {\protect\JournalTitle{Physical Review Letters}} \textbf{120}, 193904
		(2018).
		
		\bibitem{wenCompactHighperformanceVortex2020}
		Y.~Wen, I.~Chremmos, Y.~Chen, G.~Zhu, J.~Zhang, J.~Zhu, Y.~Zhang, J.~Liu, and
		S.~Yu, \enquote{Compact and high-performance vortex mode sorter for
			multi-dimensional multiplexed fiber communication systems,}
		{\protect\JournalTitle{Optica}} \textbf{7}, 254--262 (2020).
		
		\bibitem{guoSpindecoupledMetasurfaceSimultaneous2021}
		Y.~Guo, S.~Zhang, M.~Pu, Q.~He, J.~Jin, M.~Xu, Y.~Zhang, P.~Gao, and X.~Luo,
		\enquote{Spin-decoupled metasurface for simultaneous detection of spin and
			orbital angular momenta via momentum transformation,}
		{\protect\JournalTitle{Light: Science \& Applications}} \textbf{10}, 63
		(2021).
		
		\bibitem{chengUltracompactOrbitalAngular2022}
		J.~Cheng, X.~Sha, H.~Zhang, Q.~Chen, G.~Qu, Q.~Song, S.~Yu, and S.~Xiao,
		\enquote{Ultracompact orbital angular momentum sorter on a cmos chip,}
		{\protect\JournalTitle{Nano Letters}} \textbf{22}, 3993--3999 (2022).
		
		\bibitem{zhouSortingPhotonsRadial2017}
		Y.~Zhou, M.~Mirhosseini, D.~Fu, J.~Zhao, S.~M. Hashemi~Rafsanjani, A.~E.
		Willner, and R.~W. Boyd, \enquote{Sorting photons by radial quantum number,}
		{\protect\JournalTitle{Physical Review Letters}} \textbf{119}, 263602 (2017).
		
		\bibitem{guGouyPhaseRadial2018}
		X.~Gu, M.~Krenn, M.~Erhard, and A.~Zeilinger, \enquote{Gouy phase radial mode
			sorter for light: Concepts and experiments,} {\protect\JournalTitle{Physical
				Review Letters}} \textbf{120}, 103601 (2018).
		
		\bibitem{weiActiveSortingOrbital2020}
		S.~Wei, S.~K. Earl, J.~Lin, S.~S. Kou, and X.-C. Yuan, \enquote{Active sorting
			of orbital angular momentum states of light with a cascaded tunable
			resonator,} {\protect\JournalTitle{Light: Science \& Applications}}
		\textbf{9}, 10 (2020).
		
		\bibitem{zhangMimickingFaradayRotation2014}
		W.~Zhang, Q.~Qi, J.~Zhou, and L.~Chen, \enquote{Mimicking faraday rotation to
			sort the orbital angular momentum of light,} {\protect\JournalTitle{Physical
				Review Letters}} \textbf{112}, 153601 (2014).
		
		\bibitem{linAllopticalMachineLearning2018}
		X.~Lin, Y.~Rivenson, N.~T. Yardimci, M.~Veli, Y.~Luo, M.~Jarrahi, and A.~Ozcan,
		\enquote{All-optical machine learning using diffractive deep neural
			networks,} {\protect\JournalTitle{Science}} \textbf{361}, 1004--1008 (2018).
		
		\bibitem{liClassspecificDifferentialDetection2019}
		J.~Li, D.~Mengu, Y.~Luo, Y.~Rivenson, and A.~Ozcan, \enquote{Class-specific
			differential detection in diffractive optical neural networks improves
			inference accuracy,} {\protect\JournalTitle{Advanced Photonics}} \textbf{1},
		046001 (2019).
		
		\bibitem{yanFourierspaceDiffractiveDeep2019}
		T.~Yan, J.~Wu, T.~Zhou, H.~Xie, F.~Xu, J.~Fan, L.~Fang, X.~Lin, and Q.~Dai,
		\enquote{Fourier-space diffractive deep neural network,}
		{\protect\JournalTitle{Physical Review Letters}} \textbf{123}, 023901 (2019).
		
		\bibitem{menguScaleShiftRotationInvariant2021}
		D.~Mengu, Y.~Rivenson, and A.~Ozcan, \enquote{Scale-, shift-, and
			rotation-invariant diffractive optical networks,} {\protect\JournalTitle{ACS
				Photonics}} \textbf{8}, 324--334 (2021).
		
		\bibitem{zhouLargescaleNeuromorphicOptoelectronic2021}
		T.~Zhou, X.~Lin, J.~Wu, Y.~Chen, H.~Xie, Y.~Li, J.~Fan, H.~Wu, L.~Fang, and
		Q.~Dai, \enquote{Large-scale neuromorphic optoelectronic computing with a
			reconfigurable diffractive processing unit,} {\protect\JournalTitle{Nature
				Photonics}} \textbf{15}, 367--373 (2021).
		
		\bibitem{rahmanEnsembleLearningDiffractive2021}
		M.~S.~S. Rahman, J.~Li, D.~Mengu, Y.~Rivenson, and A.~Ozcan, \enquote{Ensemble
			learning of diffractive optical networks,}
		{\protect\JournalTitle{Light-Science \& Applications}} \textbf{10}, 14
		(2021).
		
		\bibitem{menguClassificationReconstructionSpatially2022}
		D.~Mengu, M.~Veli, Y.~Rivenson, and A.~Ozcan, \enquote{Classification and
			reconstruction of spatially overlapping phase images using diffractive
			optical networks,} {\protect\JournalTitle{Scientific Reports}} \textbf{12},
		8446 (2022).
		
		\bibitem{luoComputationalImagingComputer2022}
		Y.~Luo, Y.~Zhao, J.~Li, E.~{\c C}etinta{\c s}, Y.~Rivenson, M.~Jarrahi, and
		A.~Ozcan, \enquote{Computational imaging without a computer: Seeing through
			random diffusers at the speed of light,} {\protect\JournalTitle{eLight}}
		\textbf{2}, 4 (2022).
		
		\bibitem{baiImageNotImage2022}
		B.~Bai, Y.~Luo, T.~Gan, J.~Hu, Y.~Li, Y.~Zhao, D.~Mengu, M.~Jarrahi, and
		A.~Ozcan, \enquote{To image, or not to image: Class-specific diffractive
			cameras with all-optical erasure of undesired objects,}
		{\protect\JournalTitle{eLight}} \textbf{2}, 14 (2022).
		
		\bibitem{menguAllOpticalPhase2022}
		D.~Mengu and A.~Ozcan, \enquote{All-optical phase recovery: Diffractive
			computing for quantitative phase imaging,} {\protect\JournalTitle{Advanced
				Optical Materials}} \textbf{10}, 2200281 (2022).
		
		\bibitem{xuMultichannelOpticalComputing2022}
		Z.~Xu, X.~Yuan, T.~Zhou, and L.~Fang, \enquote{A multichannel optical computing
			architecture for advanced machine vision,} {\protect\JournalTitle{Light:
				Science \& Applications}} \textbf{11}, 255 (2022).
		
		\bibitem{luoDesignTaskspecificOptical2019}
		Y.~Luo, D.~Mengu, N.~T. Yardimci, Y.~Rivenson, M.~Veli, M.~Jarrahi, and
		A.~Ozcan, \enquote{Design of task-specific optical systems using broadband
			diffractive neural networks,} {\protect\JournalTitle{Light-Science \&
				Applications}} \textbf{8}, 112 (2019).
		
		\bibitem{fontaineLaguerreGaussianModeSorter2019}
		N.~K. Fontaine, R.~Ryf, H.~Chen, D.~T. Neilson, K.~Kim, and J.~Carpenter,
		\enquote{Laguerre-gaussian mode sorter,} {\protect\JournalTitle{Nature
				Communications}} \textbf{10}, 1865 (2019).
		
		\bibitem{brandtHighdimensionalQuantumGates2020}
		F.~Brandt, M.~Hiekkam{\"a}ki, F.~Bouchard, M.~Huber, and R.~Fickler,
		\enquote{High-dimensional quantum gates using full-field spatial modes of
			photons,} {\protect\JournalTitle{Optica}} \textbf{7}, 98--107 (2020).
		
		\bibitem{veliTerahertzPulseShaping2021}
		M.~Veli, D.~Mengu, N.~T. Yardimci, Y.~Luo, J.~Li, Y.~Rivenson, M.~Jarrahi, and
		A.~Ozcan, \enquote{Terahertz pulse shaping using diffractive surfaces,}
		{\protect\JournalTitle{Nature Communications}} \textbf{12}, 37 (2021).
		
		\bibitem{fangPerformanceOptimizationMultiplane2021}
		J.~Fang, J.~Bu, J.~Li, C.~Lin, A.~Kong, X.~Yin, H.~Luo, X.~Song, Z.~Xie,
		T.~Lei, and X.~Yuan, \enquote{Performance optimization of multi-plane light
			conversion (mplc) mode multiplexer by error tolerance analysis,}
		{\protect\JournalTitle{Optics Express}} \textbf{29}, 37852--37861 (2021).
		
		\bibitem{hiekkamakiHighDimensionalTwoPhotonInterference2021}
		M.~Hiekkam{\"a}ki and R.~Fickler, \enquote{High-dimensional two-photon
			interference effects in spatial modes,} {\protect\JournalTitle{Physical
				Review Letters}} \textbf{126}, 123601 (2021).
		
		\bibitem{liPolarizationMultiplexedDiffractive2022}
		J.~Li, Y.-C. Hung, O.~Kulce, D.~Mengu, and A.~Ozcan, \enquote{Polarization
			multiplexed diffractive computing: All-optical implementation of a group of
			linear transformations through a polarization-encoded diffractive network,}
		{\protect\JournalTitle{Light: Science \& Applications}} \textbf{11}, 153
		(2022).
		
		\bibitem{weiParallelArrayedWaveguide2022}
		Z.~Wei, A.~Kong, J.~Hu, T.~Lei, and X.~Yuan, \enquote{Parallel arrayed
			waveguide grating for wavelength-mode hybrid multiplexing,}
		{\protect\JournalTitle{Optics Letters}} \textbf{47}, 4311--4314 (2022).
		
		\bibitem{jiaCompensatingDistortedOAM2022}
		Q.~Jia, R.~Feng, B.~Shi, F.~Sun, Y.~Zhang, H.~Li, X.~Li, Y.~Cao, J.~Wang, and
		W.~Ding, \enquote{Compensating the distorted oam beams with near zero time
			delay,} {\protect\JournalTitle{Applied Physics Letters}} \textbf{121}, 011104
		(2022).
		
		\bibitem{luoMetasurfaceenabledOnchipMultiplexed2022}
		X.~Luo, Y.~Hu, X.~Ou, X.~Li, J.~Lai, N.~Liu, X.~Cheng, A.~Pan, and H.~Duan,
		\enquote{Metasurface-enabled on-chip multiplexed diffractive neural networks
			in the visible,} {\protect\JournalTitle{Light: Science \& Applications}}
		\textbf{11}, 1--11 (2022).
		
		\bibitem{jiaUniversalTranslationOperator2022}
		Q.~Jia, R.~Feng, B.~Shi, Y.~Zhang, H.~Li, X.~Li, F.~Sun, Y.~Cao, H.~Shi,
		J.~Wang, and W.~Ding, \enquote{Universal translation operator for
			laguerre\textendash gaussian mode sorting,} {\protect\JournalTitle{Applied
				Physics Letters}} \textbf{121}, 191104 (2022).
		
		\bibitem{goodmanIntroductionFourierOptics2005a}
		J.~W. Goodman, \emph{Introduction to Fourier Optics} (Roberts and Company
		Publishers, Englewood, 2005), 3rd ed.
		
		\bibitem{renAdaptiveopticsbasedSimultaneousPre2014}
		Y.~Ren, G.~Xie, H.~Huang, N.~Ahmed, Y.~Yan, L.~Li, C.~Bao, M.~P.~J. Lavery,
		M.~Tur, M.~A. Neifeld, R.~W. Boyd, J.~H. Shapiro, and A.~E. Willner,
		\enquote{Adaptive-optics-based simultaneous pre- and post-turbulence
			compensation of multiple orbital-angular-momentum beams in a bidirectional
			free-space optical link,} {\protect\JournalTitle{Optica}} \textbf{1},
		376--382 (2014).
		
		\bibitem{fuNonprobeCompensationOptical2017}
		S.~Fu, T.~Wang, S.~Zhang, Z.~Zhang, Y.~Zhai, and C.~Gao, \enquote{Non-probe
			compensation of optical vortices carrying orbital angular momentum,}
		{\protect\JournalTitle{Photonics Research}} \textbf{5}, 251--255 (2017).
		
	\end{thebibliography}

\end{document}